\documentclass[prl,twocolumn,floatfix,altaffilletter,tightenlines,showpacs,showkeys,
               notitlepage,preprintnumbers,nofootinbib]{revtex4-1}

\usepackage{color}
\usepackage[colorlinks=true,citecolor=purple,linkcolor=blue]{hyperref}
\usepackage[normalem]{ulem}
\usepackage{amsmath,amssymb}
\usepackage{bbm}
\usepackage{epsfig}
\usepackage{graphicx}               
\usepackage{url}
\usepackage{slashed}
\usepackage{multirow}
\usepackage{placeins}
\usepackage[dvipsnames]{xcolor}
\usepackage{epstopdf}
\usepackage[utf8]{inputenc}
\usepackage{tikz}
\usetikzlibrary{trees}
\usetikzlibrary{decorations.pathmorphing}
\usetikzlibrary{decorations.markings}

\definecolor{jblue}  {RGB}{20,50,100}
\definecolor{npurple}  {RGB} {153, 51, 204}
\definecolor{wred}   {RGB}{217,0,56}
\definecolor{white}   {RGB}{255,255,255}

\definecolor{korange}   {RGB}{235, 80,  43}
\definecolor{korange2}   {RGB}{245, 100,  63}
\definecolor{kyelloworange}   {RGB}{255, 210,  110}
\definecolor{kyelloworange2}   {RGB}{240, 170,  90}
\definecolor{kred}   {RGB}{204,  102, 153}
\definecolor{kpurple}   {RGB}{153,  61, 190}
\definecolor{kpurplelight}   {RGB}{213,  161, 230}

\usepackage{cleveref}
\usepackage{subfigure} 
\usepackage{lipsum}
\definecolor{red}{rgb}{1.0, 0, 0}
 \usepackage{gensymb}
\allowdisplaybreaks

\bibpunct{[}{]}{,}{n}{}{,}

\newcommand{\diag}{\text{diag}}

\renewcommand{\vec}[1]{{\mathbf{#1}}}


\begin{document}

\title{Fuzzy Dark Matter and Non-Standard Neutrino Interactions}
\author{Vedran Brdar}    \email{vbrdar@uni-mainz.de}
\author{Joachim Kopp}    \email{jkopp@uni-mainz.de}
\author{Jia Liu}         \email{liuj@uni-mainz.de}
\author{Pascal Prass}    \email{pprass@students.uni-mainz.de}
\author{Xiao-Ping Wang}  \email{xiaowang@uni-mainz.de}
\affiliation{PRISMA Cluster of Excellence and
             Mainz Institute for Theoretical Physics,
             Johannes Gutenberg-Universit\"{a}t Mainz, 55099 Mainz, Germany}
\date{May 24, 2017}

\preprint{MITP/17-037}

\begin{abstract}
  We discuss novel ways in which neutrino oscillation experiments can probe
  dark matter.  In particular, we focus on interactions between neutrinos
  and ultra-light (``fuzzy'') dark matter particles with masses of order $10^{-22}$\,eV.
  It has been shown previously that such dark matter candidates are
  phenomenologically successful and might help ameliorate the
  tension between predicted and observed small scale structures in the
  Universe. We argue that coherent forward scattering of neutrinos on fuzzy
  dark matter particles can significantly alter neutrino oscillation
  probabilities.  These effects could be observable in current and future
  experiments. We set new limits on fuzzy dark matter interacting with neutrinos
  using T2K and solar neutrino data, and we estimate the sensitivity of
  reactor neutrino experiments and of future long-baseline accelerator
  experiments. These results are based on detailed simulations in GLoBES.
  We allow the dark matter particle to be either a scalar or a
  vector boson.  In the latter case, we find potentially interesting
  connections to models addressing various $B$ physics anomalies.
\end{abstract}

\maketitle

Our ignorance about the particle physics nature of dark matter (DM) is so vast
that viable candidate particles span more than 90 orders of magnitude in mass.
At the heavy end of the spectrum are primordial black holes ~\cite{
  Bird:2016dcv, Carr:2016drx, Chen:2016pud, Georg:2017mqk,
Garcia-Bellido:2017fdg}.  On the low end of the DM mass spectrum are models of
``Fuzzy Dark Matter'' with a mass of order $m_\phi \sim 10^{-22}$\,eV.  The
term ``fuzzy'' refers to the huge Compton wave length $\lambda = 2\pi/m_\phi
\simeq 0.4\,\text{pc} \times (10^{-22}\,\text{eV} / m_\phi)$ of such DM particles.
Fuzzy DM has been studied mostly in the context of axions or other extremely
light scalar fields~\cite{
Hu:2000ke,            
Arias:2012az,         
Li:2013nal,           
Hui:2016ltb,          
Menci:2017nsr,        
Diez-Tejedor:2017ivd, 
Irsic:2017yje,        
Visinelli:2017imh}.   
Such DM candidates can be searched for in laboratory experiments
using cavity-based haloscopes \cite{Sikivie:1983ip, Asztalos:2009yp,
Brubaker:2016ktl}, 
helioscopes \cite{Irastorza:2011gs, Armengaud:2014gea, Arik:2011rx,
Arik:2013nya},
LC circuits \cite{Sikivie:2013laa},
atomic clocks \cite{Derevianko:2013oaa, Arvanitaki:2014faa},
atomic spectroscopy \cite{VanTilburg:2015oza} and interferometry
\cite{Geraci:2016fva}, as well as accelerometers \cite{Graham:2015ifn}
and magnetometry \cite{Budker:2013hfa, Graham:2013gfa, Kahn:2016aff}. 
Constraints on their parameter space can also be set using 
current gravitational wave detectors \cite{Branca:2016rez, Arvanitaki:2015iga,
Arvanitaki:2016fyj}.
However, ultra-light vector bosons are also conceivable fuzzy DM
candidates~\cite{Pospelov:2008jk, Nelson:2011sf, 
  Arias:2012az,Pires:2012yr, An:2014twa, Heeck:2014zfa,
  Chaudhuri:2014dla, Graham:2015rva, Dubovsky:2015cca, Redondo:2015iea,
Lee:2016ejx, Yang:2016odu, Cembranos:2016ugq}.
The tightest constraints on the mass of Fuzzy DM come from observations of large
scale structure in the Universe, and very recent studies suggest that $m_\phi >
10^{-21}$\,eV may be required~\cite{Menci:2017nsr, Irsic:2017yje,
Armengaud:2017nkf}.

Because of its macroscopic delocalization, Fuzzy DM has the potential to
resolve several puzzles related to structure formation in the Universe:  (i)
DM delocalization can explain the observed flattening of (dwarf) galaxy
rotation curves towards their center~\cite{Hu:2000ke}, which is in tension with
predictions from $N$-body simulations~\cite{Moore:1994yx, Flores:1994gz,
Navarro:1996gj} (``cusp vs.\ core problem''); (ii) the lower than expected
abundance of dwarf galaxies~\cite{Klypin:1999uc} (``missing satellites
problem'') can be understood in Fuzzy DM scenarios because of the higher
probability for tidal disruption of DM subhalos and because of the suppression
of the matter power spectrum at small scales~\cite{Marsh:2013ywa, Hui:2016ltb};
(iii) the apparent failure of many of the most massive Milky Way subhalos to
host visible dwarf galaxies~\cite{BoylanKolchin:2011de, BoylanKolchin:2011dk}
(``too big to fail problem'') is ameliorated since Fuzzy DM predicts fewer such
subhalos \cite{Marsh:2013ywa, Hui:2016ltb}; While it is conceivable that these
galactic anomalies will disappear with a more refined treatment of baryonic
physics in simulations~\cite{Governato:2012, DiCintio:2014xia, Wetzel:2016},
the possibility that DM physics plays a crucial role is far from excluded.

Our goal in the present paper is to highlight the tremendous opportunities for
probing interactions of fuzzy DM in current and future neutrino
oscillation experiments.  These opportunities exist in particular in scenarios
in which DM--neutrino interactions are flavor non-universal or flavor
violating.  In this case, even very feeble couplings between neutrinos and dark
matter are sufficient for coherent forward scattering to induce a
non-negligible potential for neutrinos, which affects neutrino oscillation
probabilities and will thus alter the expected event rates and spectra in
current and future neutrino oscillation experiments~\cite{Berlin:2016woy,
Capozzi:2017auw}. We will in particular derive
constraints from T2K and solar neutrino neutrino data, and we will determine
the sensitivities of DUNE and RENO.  Similar effects have been
considered previously in ref.~\cite{Berlin:2016woy}, where the focus has been
on anomalous temporal modulation of neutrino oscillation
probabilities.  


\emph{Dark Matter--Neutrino Interactions.}
Fuzzy DM can consist either of scalar particles $\phi$ or of vector bosons $\phi^\mu$.
In the scalar case, the relevant terms in the Lagrangian are given
by~\cite{Krnjaic:2017zlz}
\begin{align}
  \mathcal{L}_\text{scalar}
    &= \bar\nu_L^\alpha i \slashed\partial \nu_L^\alpha
     - \tfrac{1}{2} m_\nu^{\alpha\beta} \overline{(\nu_L^c)^\alpha} \nu_L^\beta
     - \tfrac{1}{2} y^{\alpha\beta} \phi \, \overline{(\nu_L^c)^\alpha} \nu_L^\beta \,,
  \label{eq:L-scalar}
\end{align}
where $\alpha$, $\beta$ are flavor indices and $y^{\alpha\beta}$ are the coupling
constants.  For vector DM, the Lagrangian is
\begin{align}
  \mathcal{L}_\text{vector}
    &= \bar\nu_L^\alpha i \slashed\partial \nu_L^\alpha
     - \tfrac{1}{2} m_\nu^{\alpha\beta} \overline{(\nu_L^c)^\alpha} \nu_L^\beta
     + g Q^{\alpha\beta} \phi^\mu \bar{\nu}_L^\alpha \gamma_\mu \nu_L^\beta \,,
  \label{eq:L-vector}
\end{align}
with the coupling constant $g$ and the charge matrix $Q^{\alpha\beta}$.
In both Lagrangians, $m_\nu$ is the effective Majorana neutrino mass matrix.
The interaction term in \cref{eq:L-scalar} can be generated in a gauge invariant way by
coupling the scalar DM particle $\phi$ to heavy right-handed neutrinos in a
seesaw scenario~\cite{Krnjaic:2017zlz}.
The interaction in \cref{eq:L-vector} could arise for instance if the DM is the
feebly coupled gauge boson corresponding to a local $L_\mu - L_\tau$ lepton
family number symmetry, defined via $Q^{ee} = 0$, $Q^{\mu\mu} = 1$,
$Q^{\tau\tau} = -1$.  Alternatively, the DM particle could couple to the SM via
mixing with a much heavier gauge boson $Z'$ with flavor non-universal
couplings.  If the $Z'$ boson has a mass of order $m_{Z'} \sim \text{TeV}$, we
expect the mixing-induced coupling $g$ in \cref{eq:L-vector} to be of order $g
\sim m_\phi / m_{Z'}$. Intriguingly, we will see below that such tiny couplings
may be within reach of neutrino oscillation experiments.  Interesting
candidates for a TeV-scale $Z'$ boson mediating interactions of ultra-light
vector DM and neutrinos include an $L_\mu - L_\tau$ gauge boson, or a new
gauge boson coupled predominantly to the second family of leptons.  The latter
possibility is of particular interest as such a particle could explain several
recent anomalies in $B$ physics~\cite{Altmannshofer:2013foa}.
We defer a detailed discussion of possible UV completions of
\cref{eq:L-scalar,eq:L-vector} to a forthcoming
publication~\cite{Brdar:inprep}.  The
mass of $\phi^\mu$ can be generated either through the St\"uckelberg mechanism
\cite{Stueckelberg:1900zz} or from spontaneous symmetry breaking in a dark
Higgs sector.


\emph{Production of Ultra-light DM Particles.}
DM particles with masses $m_\phi \ll \text{keV}$ must have been produced
non-thermally in the early Universe to avoid constraints on hot (i.e.\
relativistically moving) DM.  The most popular way to achieve this is the
misalignment mechanism, which was first introduced in the context of QCD axion
models \cite{Preskill:1982cy, Abbott:1982af, Dine:1982ah}, but can also be
applied to other ultra-light fields. For vector bosons, the misalignment
mechanism has been discussed in refs.~\cite{Nelson:2011sf, Arias:2012az,
Graham:2015rva}.  In this case, the mechanism may require also a non-minimal
coupling of $\phi^\mu$ to the Ricci scalar to avoid the need for
super-Planckian field excursions~\cite{Arias:2012az, Graham:2015rva}.  It might
be possible to avoid these extra couplings in certain UV completions of the
model~\cite{Graham:2015rva}.
The misalignment mechanism for vector bosons is more constrained
if the bosons obtain their mass through a
Higgs mechanism than in models with St\"uckelberg masses.  In particular, it
is required that the boson is massive at the temperature $T_\text{osc}$ at
which $\phi^\mu$ begins to oscillate about its minimum~\cite{Nelson:2011sf}.
This temperature is given by
$T_\text{osc} \sim \sqrt{m_\phi M_\text{Pl}}$, where $M_\text{Pl}$ is the
Planck mass.  We see that the dark Higgs boson thus needs to acquire a vacuum
expectation value (vev) $v$ at a critical temperature $T_c$ much larger than
$m_\phi \simeq g v$. Since typically $T_c \simeq
v$~\cite{Carrington:1991hz}, this implies $g \lesssim \sqrt{m_\phi / M_\text{Pl}}$.
We will, however, see that neutrino oscillation experiments are sufficiently
sensitive to probe the relevant parameter region. As an
alternative to the misalignment mechanism, the authors of
ref.~\cite{Graham:2015rva} propose production of vector DM from quantum
fluctuations during inflation, but argue that this mechanism can only account
for all the DM in the Universe if $m_\phi > 10^{-6}$\,eV.


\emph{Coherent Forward Scattering of Neutrinos on Fuzzy DM.}
By inspecting \cref{eq:L-scalar,eq:L-vector}, we observe that scalar DM
$\phi$, treated as a classical field, alters the neutrino mass
matrix, $m_\nu \to m_\nu + y \phi$, while vector DM $\phi^\mu$ 
alters their effective 4-momenta, $p_\mu \to p_\mu + g Q \phi_\mu$.
This can be seen as dynamical Lorentz violation 
\cite{Kostelecky:2003cr}. 
For implementing these effects in simulation codes, we parameterize
them in terms of a Mikheyev--Smirnov--Wolfenstein-like
potential $V_\text{eff}$~\cite{Smirnov:2003da,Wolfenstein,Mikheev:1986gs,Mikheev:1986wj}.
To do so, we use the equations of motion derived from
\cref{eq:L-scalar,eq:L-vector} (treating $\phi$ and $\phi^\mu$ as classical
fields) to derive a modified neutrino dispersion relation in the form
\begin{align}
  (E_\nu - V_\text{eff})^2 = \vec{p}_\nu^2 + m_\nu^2 \,.
  \label{eq:disp}
\end{align}
Here, $E_\nu$, $V_\text{eff}$, and $m_\nu$ should be understood as $3 \times 3$
matrices.  Neglecting the $V_\text{eff}^2$ term in \cref{eq:disp}, we read off that
\begin{align}
  V_\text{eff}
    &= \frac{1}{2E_\nu} \Big( \phi \, (y \, m_\nu + m_\nu \, y) + \phi^2 y^2 \Big) ,
    & \text{(scalar DM)}
    \label{eq:V-scalar}  \\
  V_\text{eff}
    &= -\frac{1}{2E_\nu} \Big( 2 (p_\nu \cdot \phi) g Q + g^2 Q^2 \phi^2 \Big) .
    & \text{(vector DM)}
  \label{eq:V-vector}
\end{align}
These expressions for $V_\text{eff}$ should now be added to the Hamiltonian
on which the derivation of neutrino oscillation probabilities is based.
The classical DM field can be expressed as $\phi = \phi_0 \cos(m_\phi t)$
for scalar DM and as $\phi^\mu = \phi_0 \xi^\mu \cos(m_\phi t)$ for vector DM,
where $\xi^\mu$ is a polarization vector.  The oscillation amplitude $\phi_0$
is related to the local DM energy density $\rho_\phi \sim
0.3\,\text{GeV}/\text{cm}^3$ via~\cite{Nelson:2011sf,Arias:2012az, MQED}
\begin{align}
  \phi_0 = \frac{\sqrt{2 \rho_\phi}}{m_\phi} \,.
  \label{eq:phi-0}
\end{align}
For the tiny DM masses we are interested in here, the period $\tau$ of field
oscillations is macroscopic, $\tau \simeq 1.3\,\text{yrs} \times
(10^{-22}\,\text{eV} / m_\phi)$~\cite{Stadnik:2014cea, Stadnik:2014tta,
Blas:2016ddr, Davoudiasl:2017jke}.
Note that in \cref{eq:V-scalar,eq:V-vector}, the terms linear in the couplings
constants are valid when the DM mass is so low that the DM field can be treated
as classical; the quadratic terms are approximately valid for any DM mass.

In deriving numerical results, we will for definiteness assume the neutrino--DM
couplings to have a flavor structure given by $y = y_0 (m_\nu / 0.1\,eV)$
for scalar DM, where $y_0$ is a constant
for scalar DM. This choice is motivated by the assumption of universal couplings of $\phi$
to right-handed neutrinos. For vector DM, we assume
$Q = \diag(0, 1, -1)$, as motivated by $L_\mu - L_\tau$ symmetry.  We will moreover
assume that contributions to the neutrino oscillation probabilities
proportional to powers of $\cos(m_\phi t)$ are averaged. In other words, we assume
the running time of the experiment to be much larger than $\tau$.
\Cref{eq:V-vector} shows that for vector DM, $V_\text{eff}$
depends on the polarization of the field. As it is unclear whether the initial
polarization survives structure formation or is completely randomized even on scales
$\sim 1000$\,km relevant to long-baseline experiments, we will consider both the case of
fully polarized and fully unpolarized DM. 
In the former case, we assume the
polarization axis to be parallel to the ecliptic plane for definiteness.  For
fully polarized DM, the leading contribution to $V_\text{eff}$ is linear in the
small coupling $g$, while for unpolarized DM $\xi^\mu$ varies randomly along
the neutrino trajectory, so the leading contribution to $V_\text{eff}$ is
$\mathcal{O}(g^2)$.  The same would be true for DM polarized in a direction
transverse to the neutrino trajectory.


\emph{Modified Neutrino Oscillation Probabilities.}
We have implemented the potential from \cref{eq:L-scalar,eq:L-vector} in GLoBES
\cite{Huber:2004ka, Huber:2007ji, Kopp:2006wp, Kopp:2007ne}. To facilitate
integration of the predicted event rates over time, we evaluate the oscillation
probabilities at several fixed times and interpolate them using a second
order polynomial in $\cos(m_\phi t)$.  The latter can then be integrated analytically.
We do not include long-term temporal modulation effects in our fits because the
available long-baseline data is presented in time-integrated form.
We have checked that including modulation with time in the fit does not significantly
improve our results~\cite{Brdar:inprep}.

\begin{figure}
  \centering
  \includegraphics[width=\columnwidth]{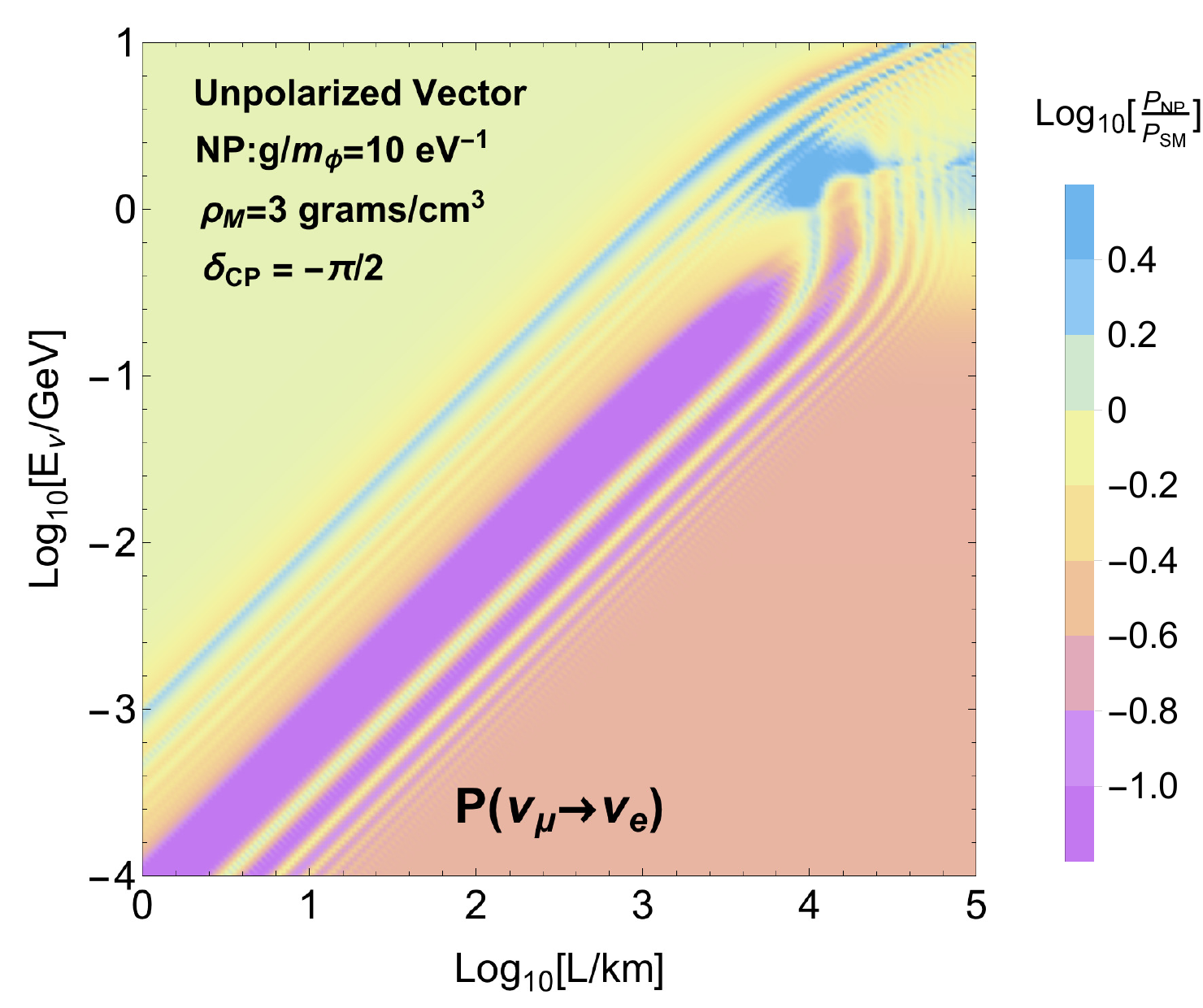}
  \caption{Impact of neutrino--DM interactions on neutrino oscillation probabilities
    in the $\nu_\mu \to \nu_e$, shown as a function of baseline $L$ and energy $E_\nu$.
    The color-code shows the ratio $P_\text{NP} / P_\text{SM}$, where
    $P_\text{NP}$ is the oscillation probability in the presence of unpolarized
    fuzzy vector DM coupled to neutrinos with $Q^{ee} = 0$, $Q^{\mu\mu} = 1$,
    $Q^{\tau\tau} = -1$, and $P_\text{SM}$ is the standard oscillation
    probability.}
  \label{fig:probLEplane}
\end{figure}

In \cref{fig:probLEplane} we show the impact of neutrino--DM interactions on the
oscillation probabilities as a function of neutrino energy $E_\nu$ and baseline
$L$. We see that even for tiny couplings, substantial modifications are
possible.

\begin{figure*}
  \centering
  \begin{tabular}{ccc}
    \includegraphics[width=.33\textwidth]{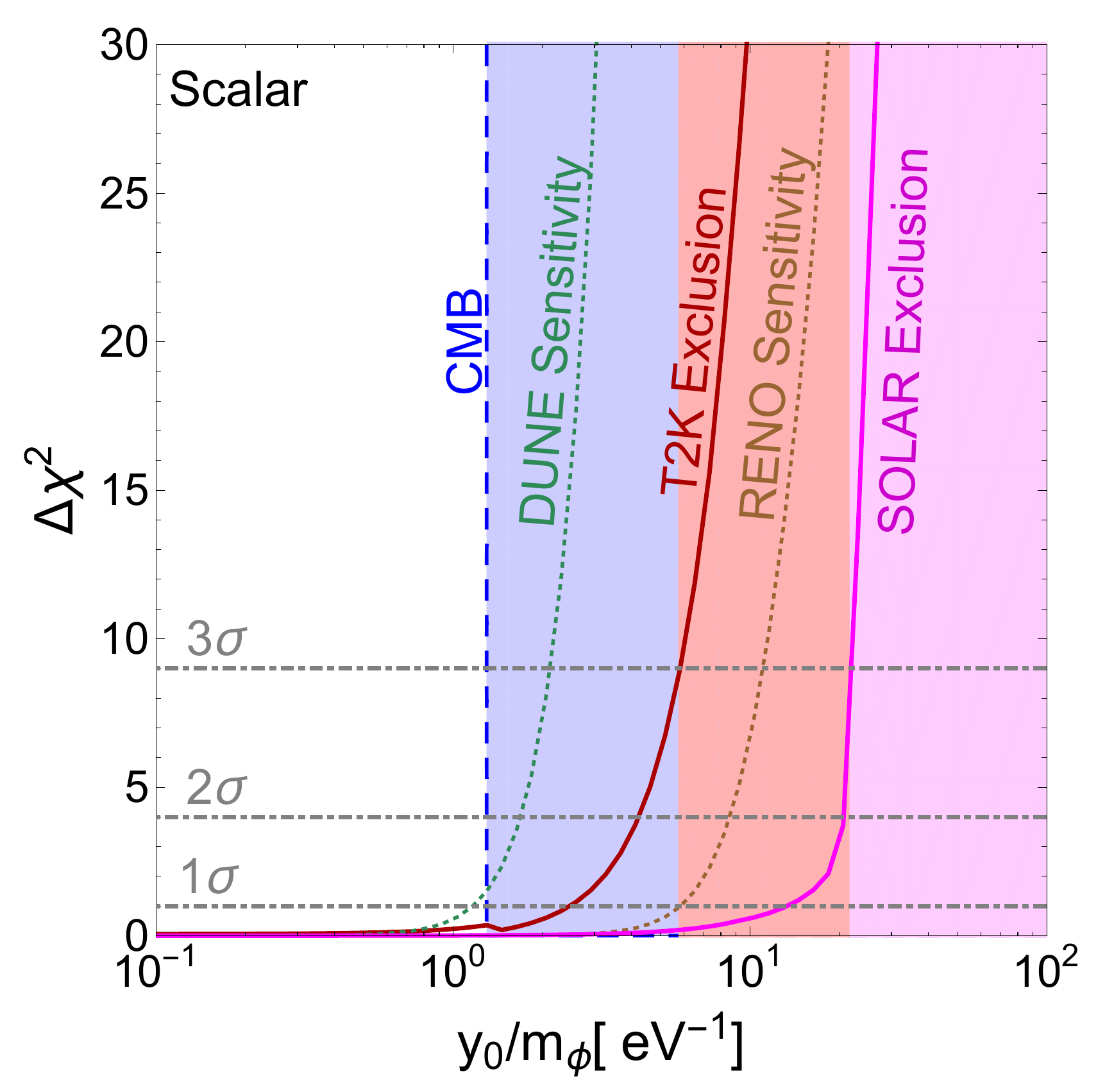} &
    \includegraphics[width=.33\textwidth]{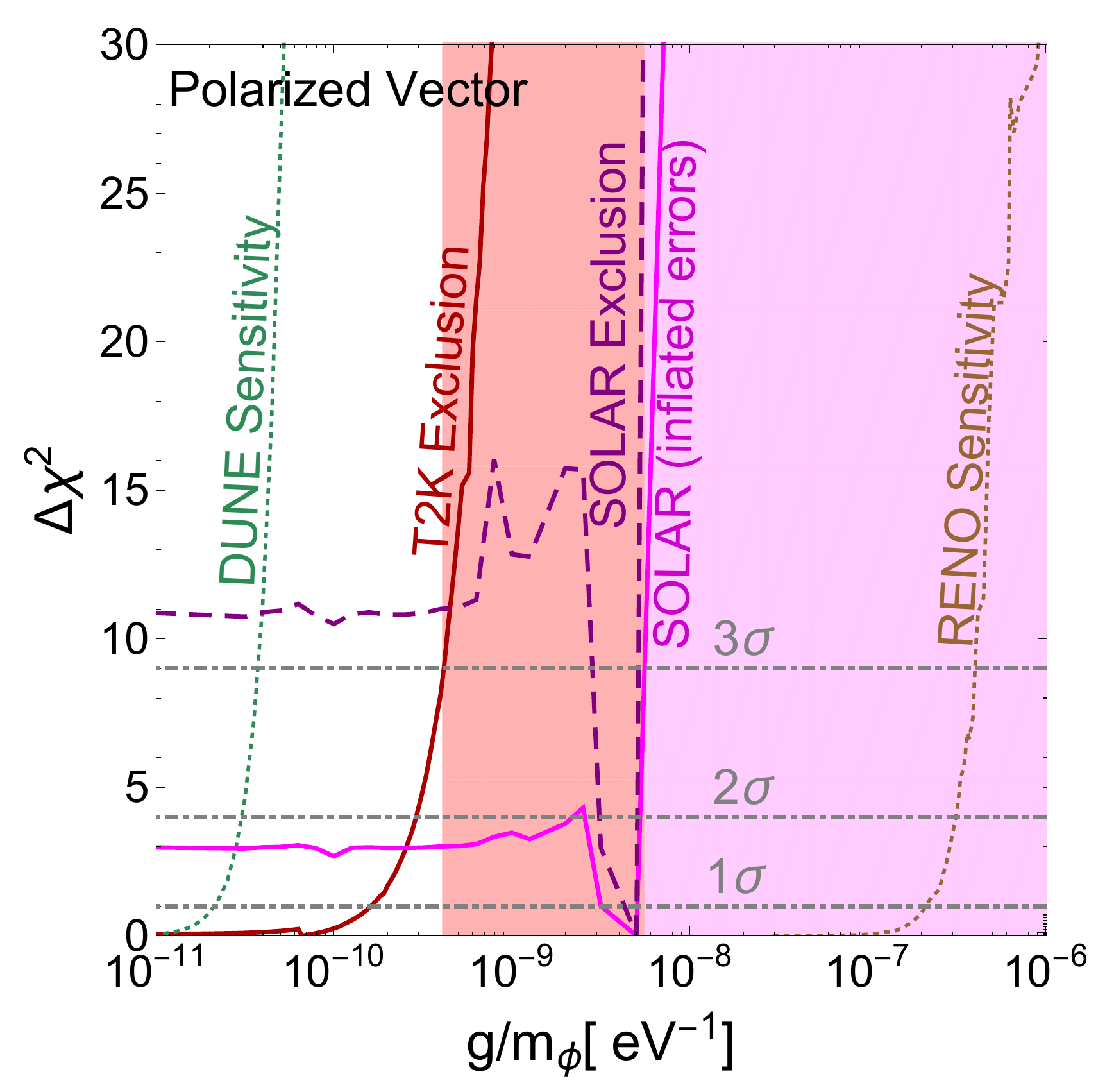} &
    \includegraphics[width=.33\textwidth]{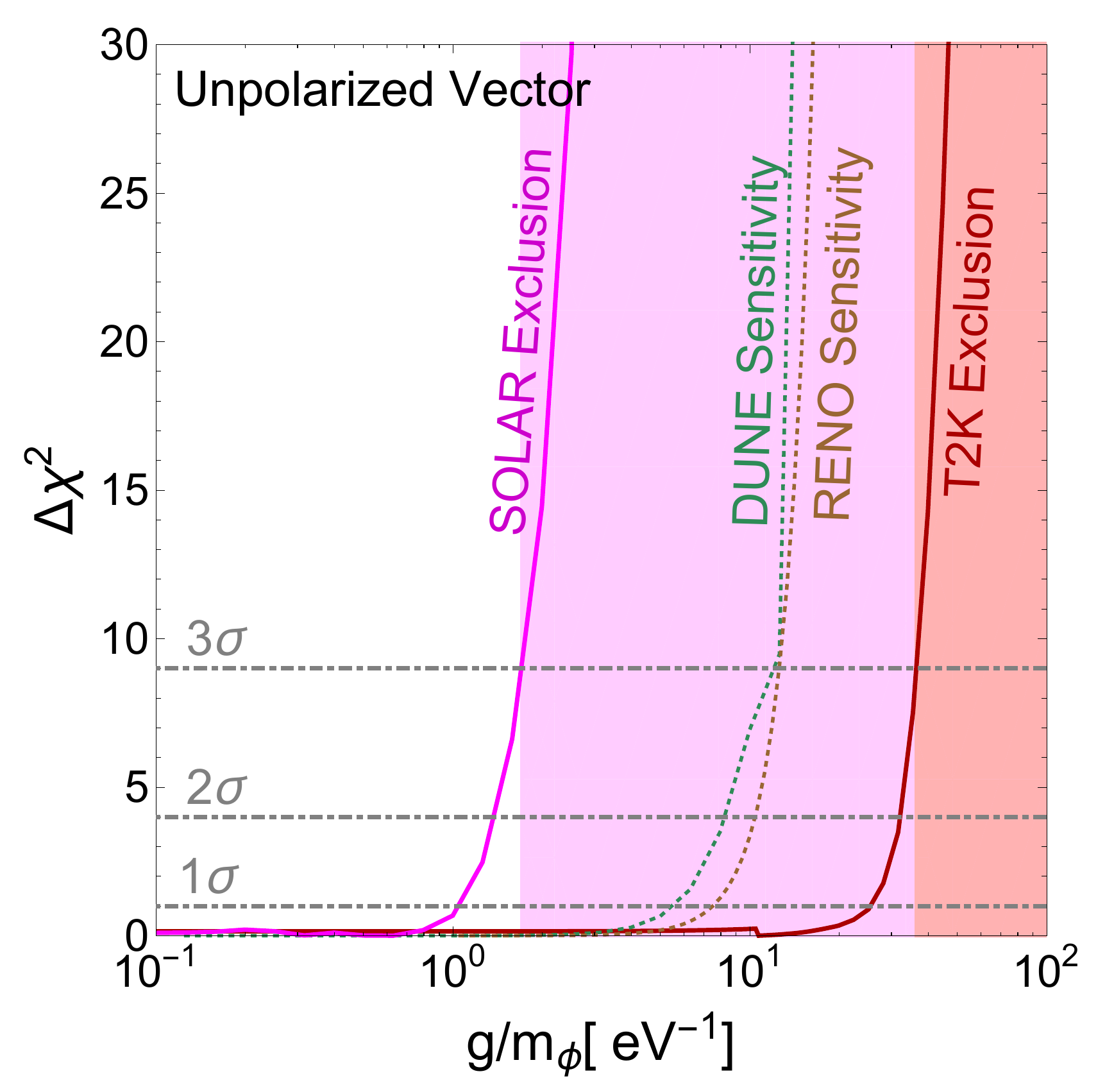} \\
    (a) & (b) & (c)
  \end{tabular}
  \caption{$\Delta\chi^2$ curves from existing (thick solid curves)
    and future (thin dotted curves) analyses of neutrino oscillation
    data.  Shaded parameter regions are excluded by the current data.
    Panel (a) applies to scalar DM, panel (b) is for vector DM with fixed
    polarization parallel to the ecliptic plane, and panel (c) is for
    unpolarized vector DM.
    For scalar DM, we have assumed $y = y_0 (m_\nu / 0.1\,eV)$,
    while for vector DM, we use a coupling structure inspired by
    $L_\mu - L_\tau$ symmetry, namely $Q^{ee} = 0$, $Q^{\mu\mu} = 1$,
    $Q^{\tau\tau} = -1$. In panel (a), we show also a limit
    based on the cosmological constraint on $\sum m_\nu$.
  }
  \label{fig:limits}
\end{figure*}

\emph{Signals in Long Baseline Experiments.}
In \cref{fig:limits}, we collect various limits and future sensitivities
on neutrino--DM interactions.
For the T2K experiment, we have developed a new GLoBES
implementation~\cite{Messier:1999kj,Paschos:2001np}, which we
use to fit data based on an integrated luminosity of $6.6 \times 10^{20}$
protons on target (pot)~\cite{Abe:2015awa, Escudero:2016}. We have verified
that we reproduce T2K's standard oscillation results to high accuracy before
setting limits on DM. For the projected sensitivity of DUNE~\cite{Acciarri:2016crz},
we use the simulation code released with ref.~\cite{Alion:2016uaj},
corresponding to $14.7 \times 10^{20}$\,pot for neutrinos and anti-neutrinos each.
To determine the sensitivity of RENO, we rely on a simulation based on
refs.~\cite{Huber:2004xh, Huber:2009xx, Vogel:1999zy} and corresponding to
3\,yrs of data taking.

We observe that experimental sensitivities are superb, thanks to the
scaling of $V_\text{eff}$ with $1/m_\phi$, see \cref{eq:V-scalar,eq:V-vector,eq:phi-0}.
For vector DM, the sensitivity is more than ten orders of magnitude better in
the polarized case (left panel of \cref{fig:limits}) than in the unpolarized
case. In the former case, the sensitivity comes from the term linear in $g$,
which is enhanced by $E_\nu / (g \phi)$ compared to the quadratic one.
In general, experiments exclude values of the coupling constant for which
$V_\text{eff}$ is much larger than the oscillation frequency $\sim m_\nu^2 / (2 E)$.
For scalar or polarized vector DM, long baseline
experiments have a significant edge over reactor experiments, while for unpolarized
DM, RENO is able to compete even with DUNE. The reason is the scaling of $V_\text{eff}$
with $1/E$ according to \cref{eq:V-scalar,eq:V-vector}.


\emph{Signals in Solar Neutrino Experiments.}
Since solar neutrinos evolve adiabatically as they propagate out of the Sun, their
survival probability in the electron flavor is given by
\begin{align}
  P_{ee}(E_\nu) = \sum_i |U_{ei}^\odot|^2 \, |U_{ei}^\oplus|^2,
  \label{eq:solarP}
\end{align}
where $U^\odot$ and $U^\oplus$ are the effective leptonic mixing matrices at the
center of the Sun and at Earth, respectively. $U^\odot$ is strongly
affected by both SM matter effects and DM--neutrino interactions, while $U^\oplus$
differs from the vacuum mixing matrix mainly through the DM term in our scenario.
We neglect Earth matter effects as their impact on our results would be
negligible~\cite{Maltoni:2015kca}.

\begin{figure}
  \centering
  \includegraphics[width=0.8\columnwidth]{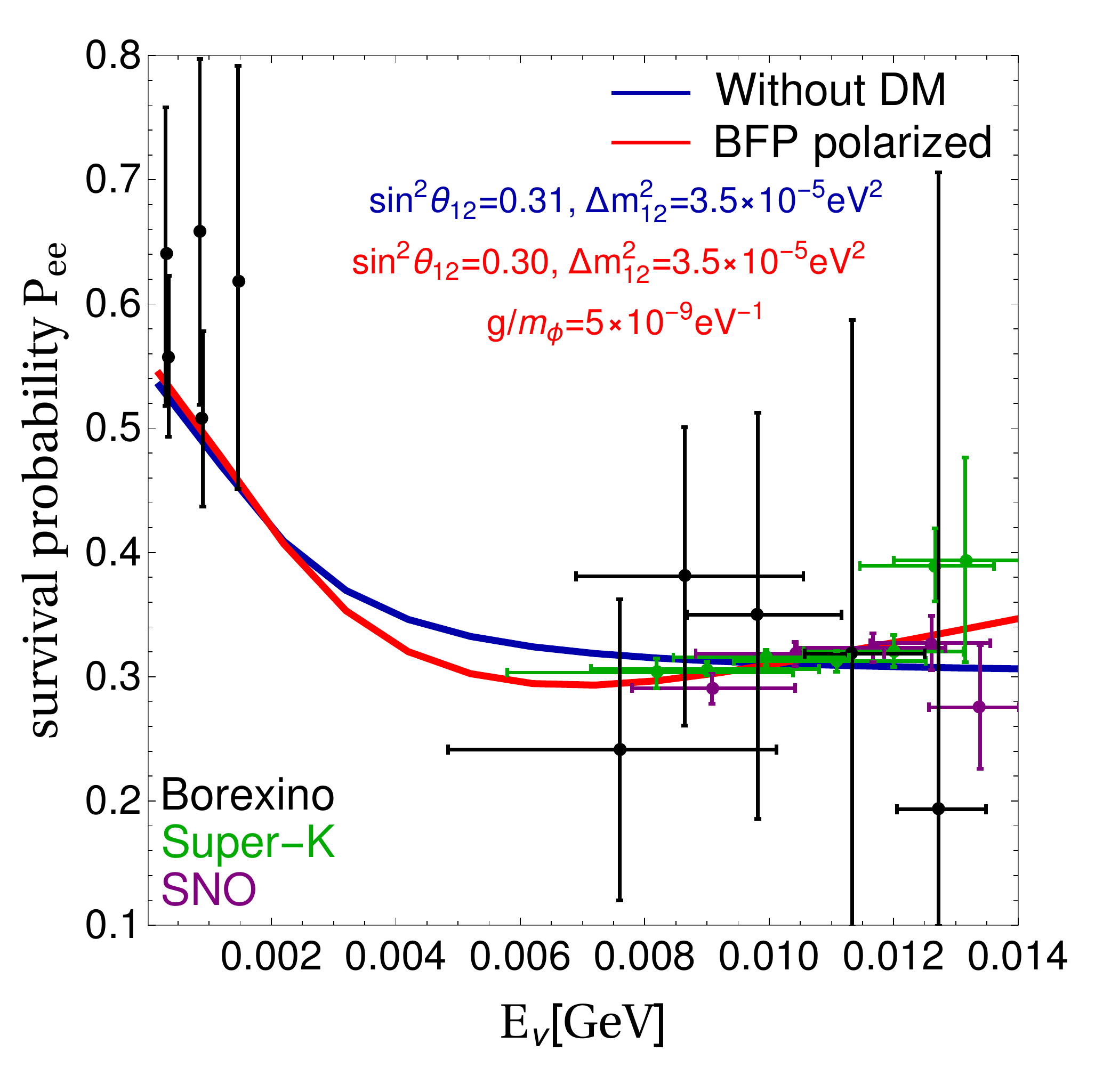} 
  \caption{Predicted $\nu_e$ survival probability for solar neutrinos at the SM
    best fit point (blue) and at the best fit point including neutrino interactions
    with polarized vector DM (red). The best fit curves for unpolarized vector DM
    and for scalar DM are similar to the SM curve. We compared to data from
    Borexino~\cite{Bellini:2008mr}, Super-Kamiokande~\cite{Renshaw:2014awa},
    and SNO~\cite{Aharmim:2009gd,Wilson:2015qga}.}
  \label{fig:solar}
\end{figure} 

We fit solar neutrino data from Borexino~\cite{Bellini:2008mr},
Super-Kamiokande~\cite{Renshaw:2014awa}, and
SNO~\cite{Aharmim:2009gd,Wilson:2015qga}, as collected in
ref.~\cite{Maltoni:2015kca}.
We illustrate in \cref{fig:solar} how the
presence of DM--neutrino interactions could improve the fit to solar neutrino
data.  For standard oscillations, we find $\chi^2 / \text{dof} \simeq 22/20$,
while the best fit point for polarized vector DM yields $\chi^2 / \text{dof}
\simeq 12/19$.  Even though we find in both cases an acceptable goodness of fit,
standard oscillations are disfavored compared to the new
physics hypothesis. This is a reflection of the fact that the upturn of the survival
probability at low energy has not been observed yet~\cite{Maltoni:2015kca}.
As the preference for new physics in our fit is somewhat stronger than
in a fit including full spectral data~\cite{Gonzalez-Garcia:2013usa},
we show in \cref{fig:limits} also conservative constraints obtained by artificially
inflating the error bars of all solar data points by a factor of two.

Comparing limits from solar neutrino observations to those from long-baseline
experiments, we see from \cref{fig:limits} that for unpolarized vector DM,
solar neutrinos offer the most powerful constraints. This is once again due to
the $1/E_\nu$ dependence of $V_\text{eff}$ in this case.  Even though the
same scaling applies to scalar DM, solar limits are much weaker because
in our benchmark scenario, neutrino--DM interactions alter only neutrino
masses (to which solar neutrinos have poor sensitivity), but not the mixing angles.
This is also the reason why the limits from ref.~\cite{Berlin:2016woy}, which
rely on variations in the mixing angle $\theta_{12}$, are not applicable here.


\emph{Cosmological Constraints on $\sum m_\nu$.}  As pointed out in
ref.~\cite{Krnjaic:2017zlz}, interactions between neutrinos and ultra-light
scalar DM are constrained by the requirement that the DM-induced contribution
to the neutrino mass term does not violate the cosmological limit on the sum of
neutrino masses, $\sum m_\nu$. We estimate this constraint in \cref{fig:limits}
(a) by requiring that, at recombination (redshift $z = 1\,100$), the correction
to the heaviest neutrino mass (taken at 0.05\,eV) should not be larger than 0.1\,eV.


\emph{Astrophysical Neutrinos.} One may wonder whether neutrino--DM
interactions could inhibit the propagation of astrophysical
neutrinos~\cite{Aartsen:2015zva} from distant sources~\cite{Arguelles:2017atb}.
The optical depth for such neutrinos is given by~\cite{Reynoso:2016hjr}
$\tau_\nu (E_\nu) = \sigma_{\nu \phi}(E_\nu) X_{\phi} m_\phi^{-1}$, with
the DM column density $X_{\phi} \equiv \int_{\text{l.o.s}} dl \, \rho_{\phi}$,
where the integral runs along the line of sight.
For both galactic and extragalactic neutrino sources, we have typically
$X_\phi \sim 10^{22}$--$10^{23}\,\text{GeV}/\text{cm}^2$~\cite{Reynoso:2016hjr}.
The scattering cross section for vector DM is approximately
\begin{align}
  \sigma_{\nu\phi}^T &\simeq 
    \frac{g^4}{8\pi} \frac{m_\nu^2}{E_{\nu}^2 m_\phi^2} \,,
    &\text{(vector DM)}
\end{align}
where the superscript $T$ indicates that, for simplicity, we have only considered
the transverse polarization states of DM. For scalar DM, the corresponding
expression is
\begin{align}
  \sigma_{\nu\phi} &\simeq \frac{g^4}{36\pi m_\nu^2} \,,
    &\text{(scalar DM)}
\end{align}
Requiring $\tau_\nu < 1$, we obtain the constraints
\begin{align}
  \frac{g}{m_\phi} &< 3 \cdot 10^8\,\text{eV}^{-1}
                        \, \bigg( \frac{E_\nu}{\text{PeV}} \bigg)^\frac{1}{2}
                           \bigg( \frac{0.1\,\text{eV}}{m_\nu} \bigg)^\frac{1}{2}
                           \bigg( \frac{10^{-22}\,\text{eV}}{m_\phi} \bigg)^\frac{1}{4} \!,
     \nonumber\\[0.2cm]
     &\hspace{4.5cm}\text{(vector DM)} \\[0.2cm]
   \frac{y}{m_\phi} &< 1.3 \cdot 10^{11}\,\text{eV}^{-1}
                        \, \bigg( \frac{m_\nu}{0.1\,\text{eV}} \bigg)^\frac{1}{2}
                           \bigg( \frac{10^{-22}\,\text{eV}}{m_\phi} \bigg)^\frac{3}{4} \!.
     \nonumber\\[0.2cm]
     &\hspace{4.5cm}\text{(scalar DM)}
\end{align}
We see that these limits are much weaker than the constraints imposed by
oscillation experiments (see \cref{fig:limits}) except for DM masses much
larger than the ones considered here and for very low neutrino energies At low
energy, however, astrophysical neutrinos cannot be observed because of
prohibitively large atmospheric backgrounds.


\emph{Summary.}
To conclude, we have demonstrated that unique opportunities exist at current
and future neutrino oscillation experiments to probe interactions between
neutrinos and ultra-light DM particles. The latter are an interesting
alternative to WIMP (Weakly Interacting Massive Particle) DM, avoiding many of
the phenomenological challenges faced by WIMPs.  A  particularly interesting
possibility, which we plan to explore further in an upcoming
publication~\cite{Brdar:inprep}, is a possible connection to flavor
non-universal new physics at the TeV scale,  as motivated by recent anomalies
in quark flavor physics.


\emph{Note added.}
While we were finalizing this paper, ref.~\cite{Krnjaic:2017zlz} appeared on
the arXiv, addressing similar questions.  While the main focus of
ref.~\cite{Krnjaic:2017zlz} (and also of the earlier
ref.~\cite{Berlin:2016woy}) is on scalar DM, we consider also DM in the form of
ultra-light gauge bosons. The authors of ref.~\cite{Krnjaic:2017zlz} have
considered a larger range of experiments for setting limits than us, while our
results are based on more detailed numerical simulations of the few most
relevant experiments.  Where our results are comparable to those of
ref.~\cite{Krnjaic:2017zlz}, they are in good agreement.


\emph{Acknowledgments.}
We would like to thank Pedro Machado, Georg Raffelt, and Felix Yu for very
helpful discussions.  This work has been funded by the German Research
Foundation (DFG) under Grant Nos.\ EXC-1098, \mbox{KO~4820/1--1}, FOR~2239,
GRK~1581, and by the European Research Council (ERC) under the European
Union's Horizon 2020 research and innovation programme (grant agreement No.\
637506, ``$\nu$Directions'').


\bibliographystyle{JHEP}
\bibliography{refs}

\end{document}